\documentclass[preprint,review,12pt]{elsarticle}
\usepackage{amssymb}
\usepackage{float}
\usepackage{hyperref}
\hypersetup{
    colorlinks=true,
    linkcolor=blue,
    filecolor=magenta,      
    urlcolor=cyan,
}
\usepackage[flushleft]{threeparttable}
\usepackage{tabu}
\usepackage{longtable}
\usepackage{booktabs}
\usepackage[singlelinecheck=false]{caption} 
\usepackage{lineno}
\usepackage[dvipsnames]{xcolor}

\journal{Measurement}

\begin{document}
\begin{frontmatter}

\title{Long-term quantification and characterisation of wind farm noise amplitude modulation}

\author[label1]{Phuc D. Nguyen \corref{cor1}}
\address[label1]{College of Science and Engineering, Flinders University, SA 5042, Australia}
\address[label2]{Adelaide Institute for Sleep Health, Flinders University, SA 5042, Australia}
\address[label3]{School of Mechanical Engineering, University of Adelaide, SA 50045, Australia}
\cortext[cor1]{ducphuc.nguyen@flinders.edu.au}
\author[label1]{Kristy L. Hansen}
\author[label2]{Peter Catcheside}
\author[label3]{Colin Hansen}
\author[label2]{Branko Zajamsek}

\begin{abstract}
The large-scale expansion of wind farms has prompted community debate regarding adverse impacts of wind farm noise (WFN). One of the most annoying and potentially sleep disturbing components of WFN is amplitude modulation (AM). Here we quantified and characterised AM over one year using acoustical and meteorological data measured at three locations near three wind farms. We found that the diurnal variation of outdoor AM prevalence was substantial, the nighttime prevalence was approximately 2 to 5 times higher than the daytime prevalence. On average, indoor AM occurred during the nighttime from 1.1 to 1.7 times less often than outdoor AM, but the indoor AM depth was higher than that measured outdoors. We observed an association between AM prevalence and sunset and sunrise. AM occurred more often at downwind and crosswind conditions. These findings provide important insights into long term WFN characteristics that will help to inform future WFN assessment guidelines. 
\end{abstract}


\end{frontmatter}


\section{INTRODUCTION}

Wind energy is one of the fastest-growing renewable energy sectors in the world \citep{gwec2019global}, reaching approximately 870 GW in 2021 (ref. \citep{windpower2021}. Despite the benefits of wind energy, some concerns remain regarding social \citep{wolsink2007wind,krohn1999public,kitzing2020multifaceted}, ecological \citep{thaker2018wind,schuster2015consolidating} and environmental impacts \citep{vautard2014regional,zhou2012impacts}. The noise generated by wind turbines is a recurring source of complaints regarding annoyance and potential sleep disturbance from residents living near wind farms \citep{Micic2018,liebich2020systematic}. Wind farm noise (WFN) contains unusual spectral and time-varying features that may exacerbate annoyance \citep{perkins2016review} and increase loudness \citep{jurado2019effect}, including infrasound, a low-frequency dominated spectrum \citep{ingielewicz2014infrasound,Zajamsek2016}, tonality \citep{liu2012tonality} and amplitude modulation (AM), which is a periodic variation of the noise level primarily related to blade rotational effects \citep{Hansen2017}. Wind farm AM is commonly described as `swish swoosh' or `rumble' and is of particular research interest due to its propensity to contribute to annoyance \citep{Lee2011,Schaffer2016,Ioannidou2016} and possible sleep disturbance \citep{smith2020laboratory}. However, its characteristics such as depth (or degree), duration, consistency and occurrence time could vary between wind farms \citep{zagubien2018analysis}.

Previous long-term WFN measurements found wind farm AM to be associated with wind direction \citep{conrady2020amplitude, Larsson2014, paulraj2017effect}, sound speed gradient, solar elevation angle, turbulence intensity \citep{Larsson2014}, and diurnal meteorological variations \citep{hansen2019prevalence,conrady2020amplitude}. The majority of these studies were carried out in cold climates where ground cover with snow during winter months and other climactic effects are clearly different from warmer climates without snow. Snow covered ground has a very high sound absorption coefficient, even at very low frequencies, and thus attenuates noise much more effectively than other ground surface types \citep{Hansen2017,bies2017engineering,ostashev2015acoustics}. Previous long-term studies \citep{conrady2020amplitude, Larsson2014} recorded only low time and frequency resolutions of acoustic data such as 1/3-octave bands or fast time-weighted SPLs which limited analyses to conventional AM detection methods \citep{Larsson2014,Bass2016} unable to reliably detect AM. Long-term quantification of AM has been predominantly carried out at distances of 1 km or less from wind farms, where WFN is dominated by mid to high frequencies ($>200$ Hz). At larger wind farm setback distances, much more typical for Australia, AM is dominated by lower frequencies ($<200$ Hz) \citep{hansen2019prevalence}. However, to date, low-frequency AM has not been systematically studied over a long period of time. Furthermore, although indoor WFN noise character is much more relevant to human perception, annoyance and sleep disturbance than outdoor levels, long-term characterisation and quantification of indoor AM has not been attempted to date, especially at long-range distances to wind farm. 

The purpose of this study was to quantify and characterise AM, and to examine relationships between AM, meteorological conditions and wind farm operational data over one year. To detect AM, we used a previously developed AM detection method based on machine learning (Authors, 2020). This allowed for accurate and reliable detection of AM in three long-term data acoustic sets measured near South Australian wind farms at locations up to 3.5 km away from the nearest wind turbine.  

\section{METHODS}
This study was approved by the Authors University Social and Behavioural Research Ethics Committee (SBREC project 7536). Residents living in the houses where measurements were conducted who provided voluntary informed written consent and received a small reimbursement for study involvement.

\subsection{Study region}

Measurements were conducted in the mid-north region of South Australia (\autoref{fig:method1}a and Supplementary Fig. S1), which has a Mediterranean climate with relatively mild winters and hot dry summers (See Supplementary Fig. S2). Noise was measured both outdoors and indoors at three residential houses (H1, H2 and H3) located between approximately 1 and 3.5 km from the nearest wind turbines of three wind farms. The selected wind farms included one with nearby turbines positioned at similar elevation levels as the residence (Wind farm 1,
\autoref{fig:method1}a), and two with all turbines positioned along the top of ridges (Wind farms 2 and 3, \autoref{fig:method1}a). The average height difference between each ridge and residence was 70 m for Wind farm 1 and 110 m for Wind farm 3. The chosen wind farms layouts, turbine types and total power capacity are presented in \autoref{fig:method1}a and \autoref{tab:tab1}. 

\begin{table}[ht!]
    \begin{threeparttable}
        \caption{Characteristics of wind farms.}
        \label{tab:tab1}
        \begin{center}
        \begin{tabular}{ p{4.5 cm} p{2.0 cm} p{2.0cm} p{2.0 cm} }
        \toprule
         Name & Hornsdale (Wind farm 1) & Hallett (Wind farm 2) & Waterloo (Wind farm 3)  \\ 
         \midrule
         Nominal capacity (MW)  & 315 & 148 & 131 \\
         Turbine size (MW)      & 3.2 & 2.1 & (3.0 \& 3.3) \\
         Operational month and year* & Jul 2016 & 2012 & Oct 2010  \\
         Wind farm latitude   & -33.058 & -33.367 & -33.983    \\
         Wind farm longitude  & 138.544 & 138.728 & 138.900    \\
         Annual output
         (mean $\pm$ s.d.) & 125 $\pm$ 97 & 54 $\pm$ 44 & 45 $\pm$ 38    \\
         \bottomrule 
        \end{tabular}
        \smallskip\scriptsize
        \begin{tablenotes}[flushleft]
         \item[]* \footnotesize Hornsdale and Hallett wind farms were developed in several phases. Above operational dates are based on a fully operational status, which is consistent with the number of wind turbines as shown in \autoref{fig:method1}a. (See Supplementary Fig. S3 for distribution of wind farm power output between Jun 2018 and Sep 2019.)
        \end{tablenotes}
        \end{center}
    \end{threeparttable}
\end{table}

\subsection{Experimental design}

\begin{figure}[ht]
\begin{center}
\includegraphics[width=14 cm]{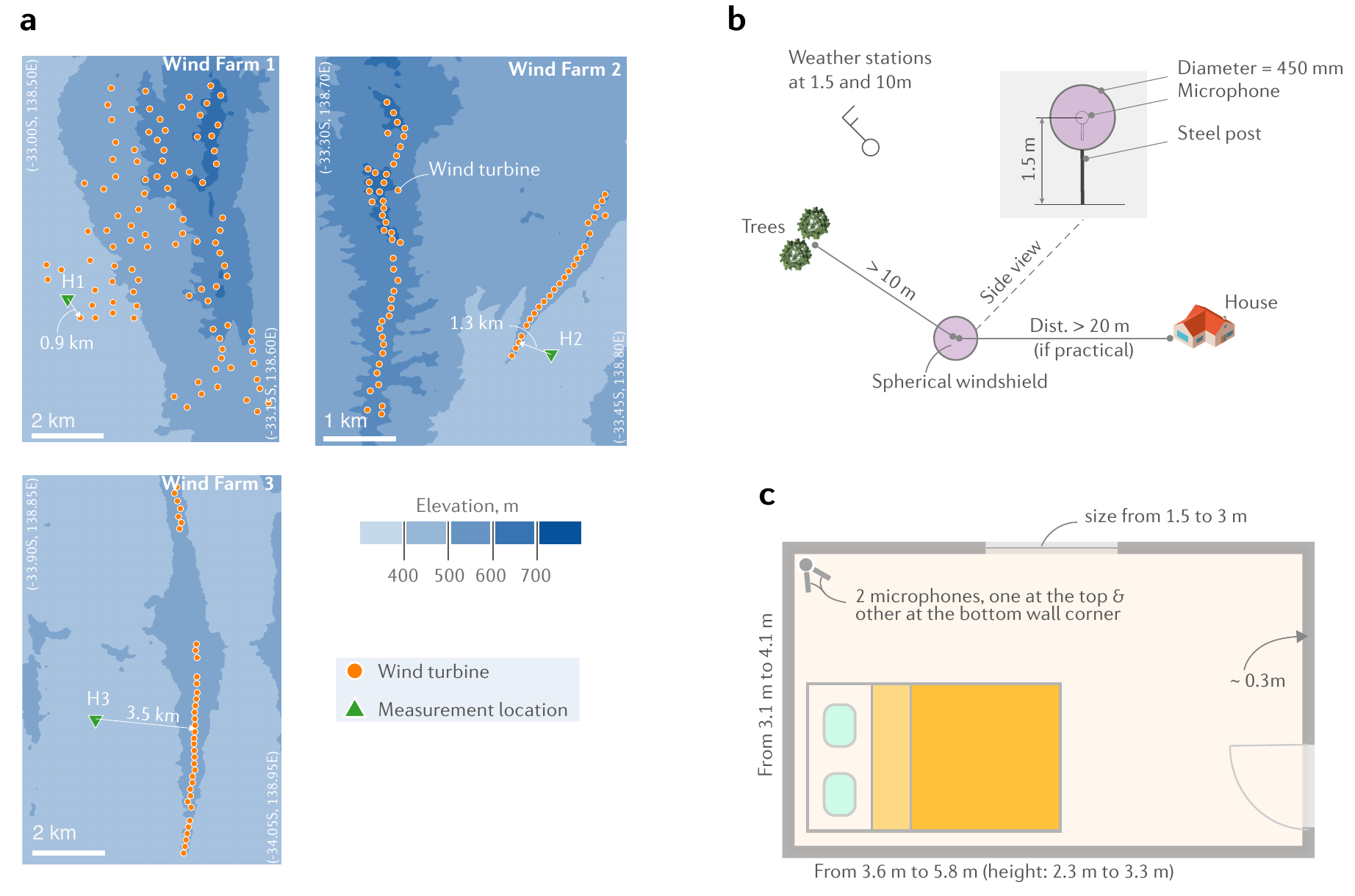}
\end{center}
\caption{Measurement locations and experimental set-up. (\textbf{a}) Wind farm layouts and measurement locations. (\textbf{b}) and (\textbf{c}) Typical outdoor and indoor microphone position set-up. }
\label{fig:method1}
\end{figure}

WFN noise was measured for more than one year, from May 2018 to August 2019 using Bruel \& Kajer LAN-XI Type 3050 data acquisition systems sampled at 8,192 Hz. The outdoor noise measurement systems consisted of two low-frequency microphones G.R.A.S type 40AZ with a noise floor of 17 dBA and flat frequency response down to 0.5 Hz, located at ground level and 1.5 m above ground (\autoref{fig:method1}b and Supplementary Fig. S4). Indoor noise was measured at the top and bottom wall corners (\autoref{fig:method1}c). The room dimensions and constructions were showed in \autoref{tab:table22}. Indoor noise levels in rural area are normally low \cite{hansen2014comp}, and thus, low-noise microphones B\&K type 4955 with a noise floor of 6.5 dB(A) and flat frequency response down to 6 Hz were used for indoor measurements. To minimise wind-induced noise, all outdoor microphones were equipped with primary and secondary windshields (see (Authors, 2014) and Supplementary Fig. S4 for design details). At location H1, the outdoor data measured at a height of 1.5 m were not available, and thus data measured at ground level were used. Although the noise measured using a microphone mounted at ground level and 1.5 m is not exactly the same \citep{hansen2014identification}, the use of 1.5 m data was still reliable, particularly for AM quantification. The agreement between AM detection results in both data sets was high (accuracy = 0.82, F1-score = 0.78, AUC = 0.9) (See Supplementary Fig. S5 for details). The AM prevalence quantified using data measured at ground level may be higher compared with 1.5 m height (See Supplementary Table S1) most likely due to lower wind-induced noise at ground level.

Local wind speed and direction were measured concurrently at 1.5 and 10 m using Davis Vantage Vue and Davis Vantage Pro weather stations. Although a general relationship between local wind speed and wind farm power output was observed (See supplementary Fig. S6), this relationship is highly uncertain. The wind speed and direction accuracy of these weather stations is 0.4 m/s and 22.5 degrees, respectively. The coarse resolution of wind direction was considered adequate to determine if the receiver was in a downwind, crosswind or upwind direction from the wind farm. To mount the weather stations at 10 m, a low cost TV antenna mast was used, slightly modified to suit the weather station and to facilitate fast set-up (see Supplementary Fig. S3). 

In addition to acoustic and meteorological data, wind farm power output capacities and digital elevation data were obtained from online data repositories (See Section 2.5).

\begin{table}[tb]
\centering
\caption{\label{tab:table22} Residential house dimensions and constructions at three measurement locations.}

\begin{tabular}{p{2cm} p{3.5cm} p{7cm} }
\toprule
Location & Room dimension* & Construction\\
\midrule
WF1   & $3.7\times5.8\times2.3$ & Timber walls; single glazed windows; pink batts ceiling insulation; concrete floor.\\
WF2              & $3.1\times3.6\times3.3$ & Red brick walls (300 mm); single glazed windows (installed double glazed windows during the measurement); pink batts ceiling insulation; wooden floor.  \\
WF3             & $4.1\times4.3\times3.2$ & Thick stone/cement brick walls (350 mm); small-medium single-pane wood-framed sash design windows; corrugated sheet steel roof; plaster panel ceiling.   \\
\bottomrule
\end{tabular}

\begin{tablenotes}[flushleft]
         \item[]* \footnotesize Dimension is length $\times$ width $\times$ height in m.
\end{tablenotes}
        
\end{table}

\subsection{Amplitude modulation detection}
Our machine learning-based random forest method was used for detecting AM and was validated using a benchmark human-scored data set (Authors, 2020). In brief, for validation purposes, an acoustic engineer listened to 6,000 10-sec audio samples randomly extracted from measured data and manually classified them as either `containing AM' versus `no AM'. 

\begin{figure}[ht]
\begin{center}
\includegraphics[width=12 cm]{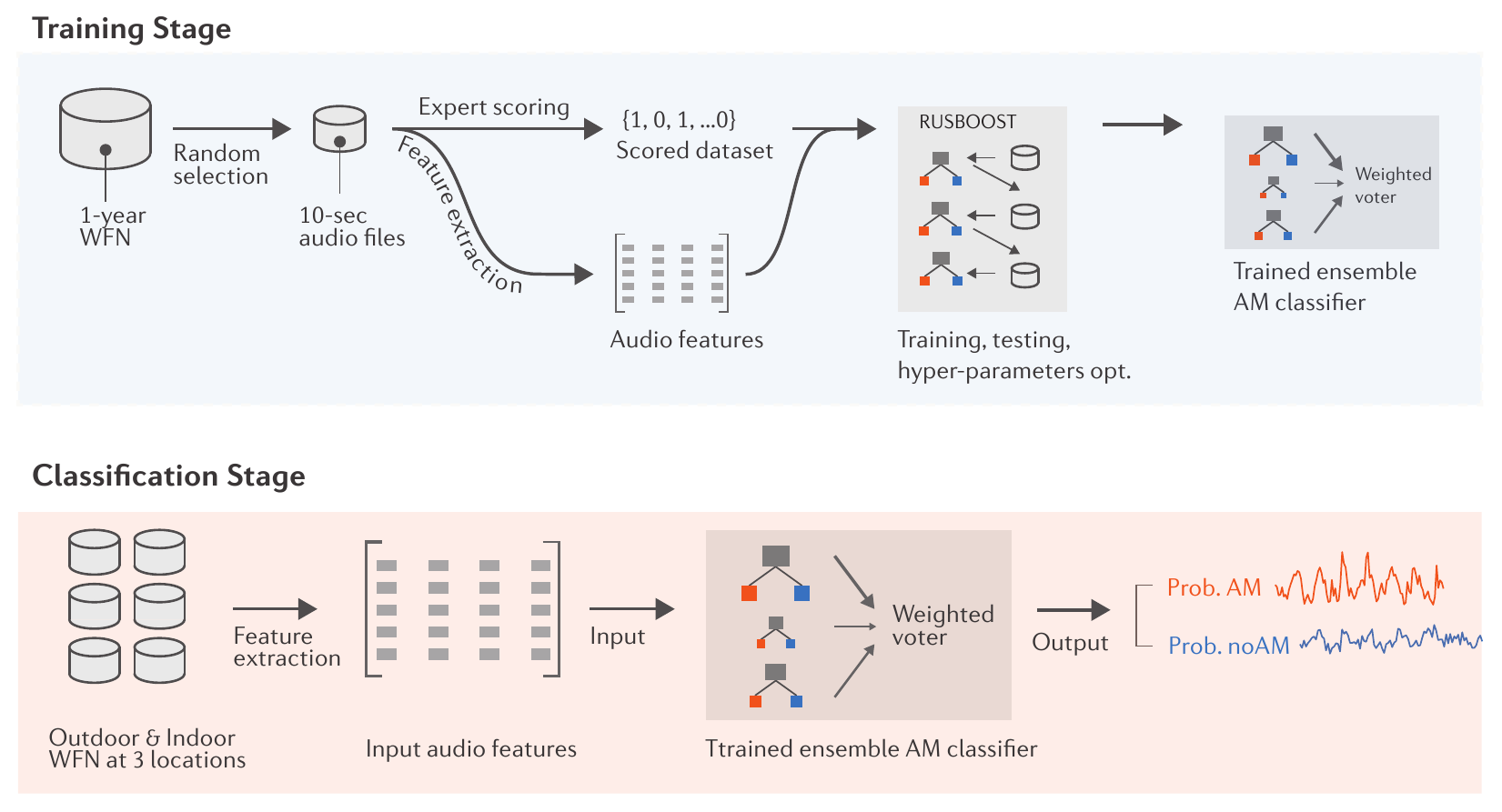}
\end{center}

\caption{Flow chart of AM detection method.}
\label{fig:flowchart}
\end{figure}

Due to the imbalance between noise samples containing AM and no AM in the data sets of the current study ($\approx20$\% AM versus $\approx80$\% no AM samples), a Random Undersampling Boosting (RUSBoost) classifier \cite{seiffert2009rusboost}, a simpler and faster alternative to SMOTEBoost \citep{chawla2003smoteboost}, was used to improve the previous developed method. Also, to maximise the performance of AM detection, a separate classifier was used for each data set (three classifiers for outdoor and three for indoor data sets). A schematic overview of the AM detection method is shown in (\autoref{fig:flowchart}). This machine learning approach showed high performance with $F$-1 scores from 0.64 to 0.8 and Matthew correlation coefficients from 0.59 to 0.78 (See Supplementary Table S2 and Fig. S7) and higher accuracy than previous methods \cite{Bass2016,Larsson2014,yokoyama2013study} (see (Authors,2020) for comparison details). 

\subsection{Data and statistical analysis}

The present study analysed the outdoor WFN noise data measured at 1.5 m above ground level (except at wind farm 1 where noise was measured at ground level) and indoor data measured in a top room corner. To ensure data quality of outdoor and indoor WFN measurements, a plot of the L$_{Aeq}$ of all data against time was constructed, and extraneous noise events were detected visually and manually excluded if noise contamination was confirmed through listening to the file. Less than 10\% of the total measured samples were excluded (See Supplementary Table S3, Fig. S8 and Fig. S9). All signal processing and data analysis were implemented in MATLAB (\url{https://www.mathworks.com}), while statistical analysis (two-tailed $t$-test and linear regression as appropriate) were implemented in R version 4.0.0 (\url{https://www.r-project.org}). The statistical significance threshold was set at $P<$ 0.05.

\subsection{Data availability}

Data to support the findings of this study are based on publicly available data including wind farm power output capacity data for each 5-minute interval accessed via \url{https://anero.id/energy/wind-energy/}, Digital Elevation Data (DEM) extracted from \url{http://www.ga.gov.au/scientific-topics/national-location-information/digital-elevation-data} and AM detection algorithms are available at (Authors). Long-term acoustical data are available upon written request to (Authors)

\section{RESULTS}

\subsection{Amplitude modulation characteristics}

\begin{figure}[ht]
\begin{center}
\includegraphics[width=12 cm]{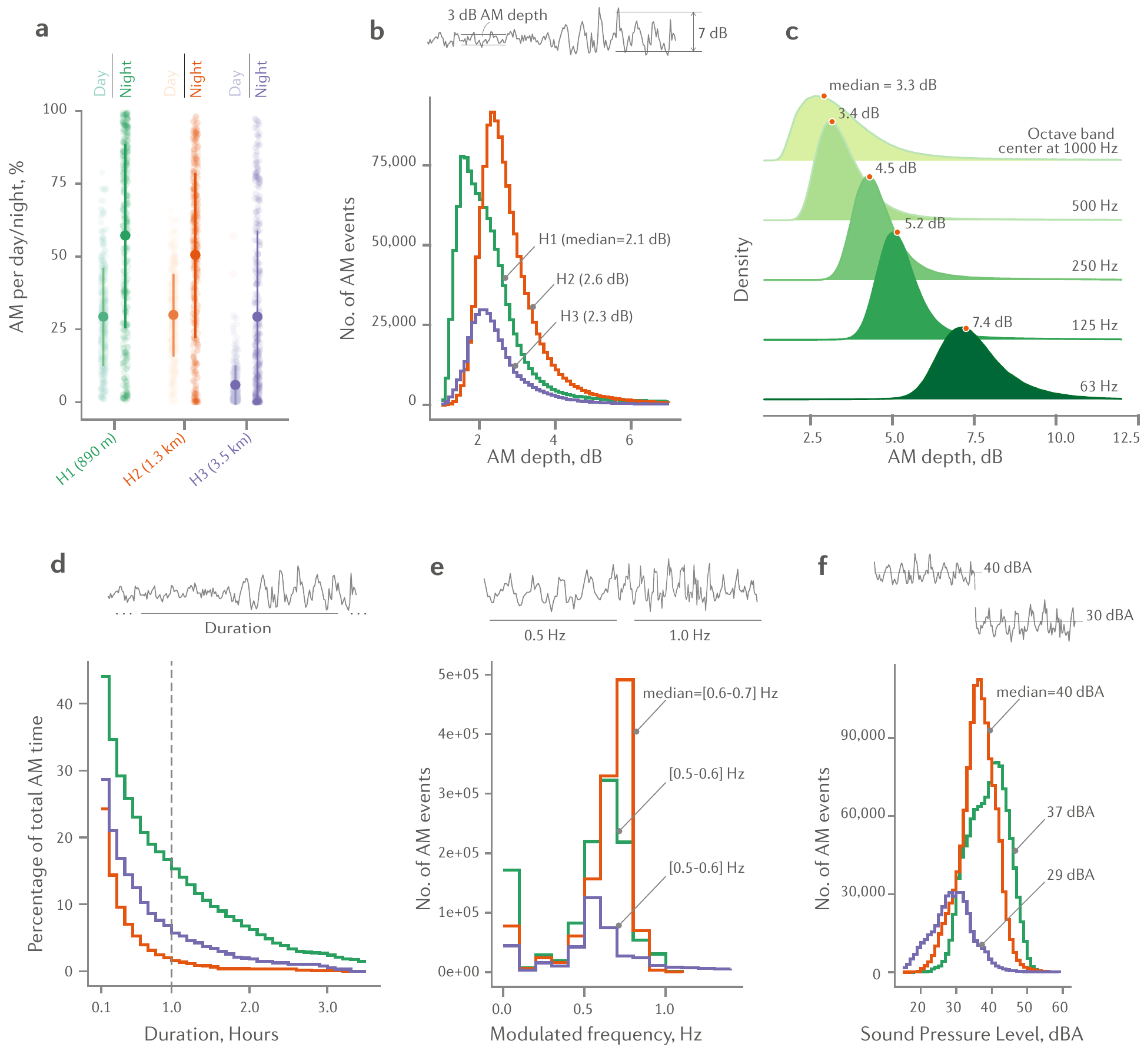}
\end{center}

\caption{Characterisation of AM for outdoor data measured over one year. (\textbf{a}) The percentage of AM during the daytime (from 7:00 to 22:00) and nighttime (22:00 to 7:00). Bold circles and error bars represent the mean and s.d. (\textbf{b}) AM depth calculated as the difference between statistical noise levels L5 and L95. (\textbf{c}) AM depth quantified in each octave band for measurements at H1. (\textbf{d}) AM duration measured as the continuous occurrence of AM events in consecutive and uninterrupted 10 s blocks. (\textbf{e}) AM frequency. (\textbf{f}) A-weighted sound pressure level associated with AM events. Examples of relevant AM characteristics in the time domain are provided above each histogram (\textbf{b,d,e \& f}).}
\label{fig:results1}
\end{figure}

AM occurred more often during the nighttime compared to the daytime (\autoref{fig:results1}a, two-sample $t$-test, all $P$-values $<$ 0.001). At locations H1 and H2 which were within 1.3 km of the nearest wind turbine, AM occurred on average for more than 50\% and 25\% of the nighttime and daytime, respectively. Similar trends were also observed at location H3, but with a lower prevalence of around 25\% AM during the nighttime and only 3\% during the daytime, where the nighttime value is comparable to previous observations for similar distances \cite{hansen2019prevalence}. The AM depth, which is a measure of the peak-to-trough variation in the overall SPL, varied between measurement locations (\autoref{fig:results1}b). The AM depth was calculated as the difference between statistical noise levels L$_{5th}$ and L$_{95th}$ of the fast time-weighted and frequency A-weighted SPL. This metric is reported in this study because it is commonly used in laboratory listening experiments assessing annoyance potential of AM \citep{yokoyama2013study,von2013wind}. The AM depth of WFN was inversely related to frequency at H1, as shown in \autoref{fig:results1}c. The results for other locations show similar trends and are provided in Supplementary Fig. S10. 

AM was an intermittent phenomenon as fewer than 20\% of consecutive 10-second AM events spanned more than one hour, as shown in \autoref{fig:results1}d. Furthermore, \autoref{fig:results1}d shows that a larger number of consecutive 10-second AM events were observed at closer locations to the wind farm, equivalent to AM lasting between 1 and 3 hours. The modulation frequency was consistently between 0.5 and 1 Hz (\autoref{fig:results1}e), which is as expected for modern wind turbines which rotate at a speed of 10 to 20 revolutions per minute \cite{Hansen2017}. However, a large number of AM events did not show a clear periodic variation (modulation frequencies between 0 and 0.5 Hz). Reasons for this may be (1) a false positive detection of AM or (2) an intermittent, rather than periodic variation in SPL. AM events were dominant at particular SPLs (median values of 39.4, 36.4 and 29.1 dBA for H1-H3, respectively) (\autoref{fig:results1}f).   

\subsection{Outdoor-to-indoor variability}

\begin{figure}[ht]
\begin{center}
\includegraphics[width=15 cm]{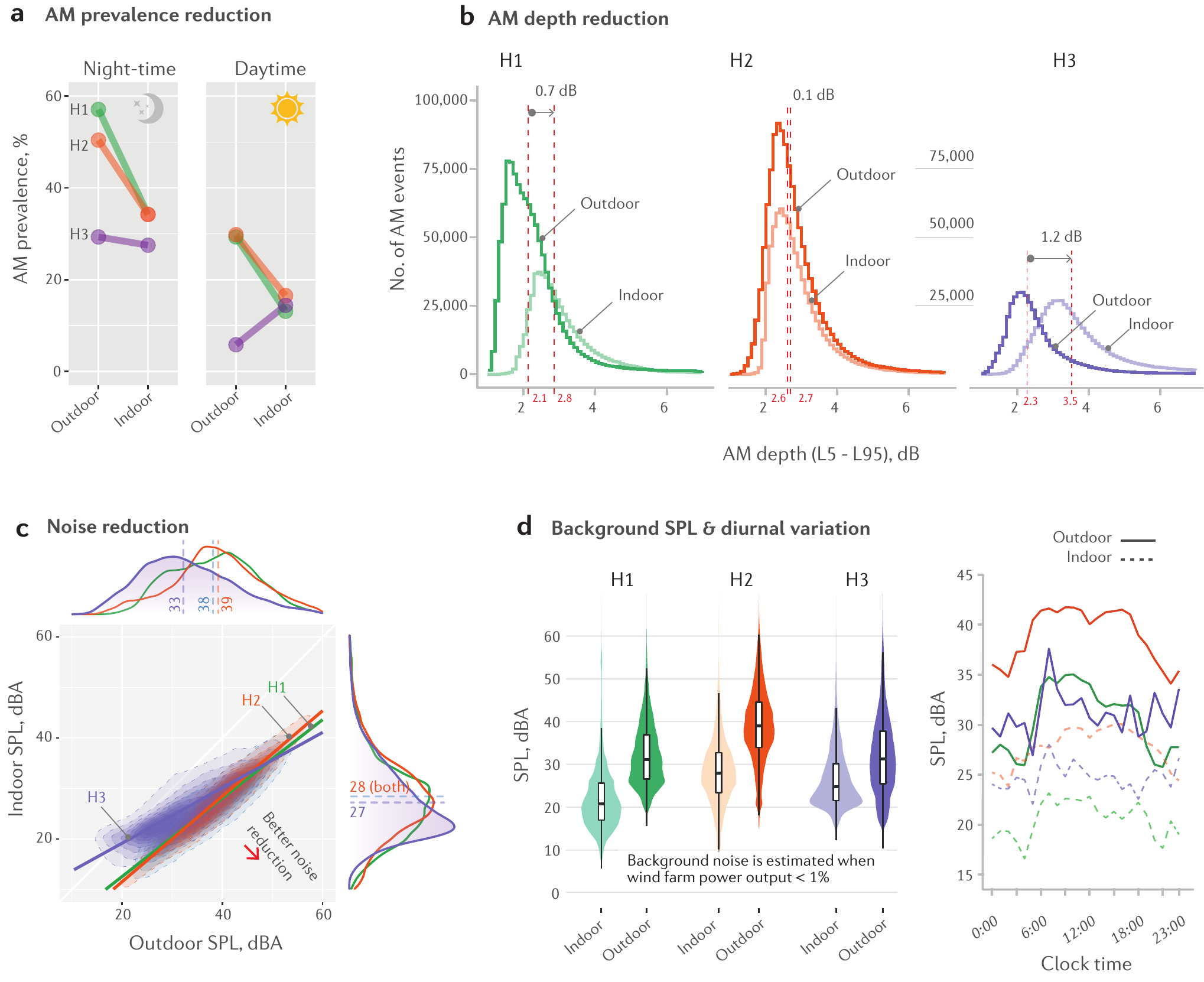}
\end{center}
\caption{Outdoor and indoor AM prevalence and AM depth. (\textbf{a}) AM prevalence percentage difference during the nighttime and daytime. (\textbf{b}) AM depth distributions measured indoors and outdoors at three locations with median values indicated by vertical red dashed lines. (\textbf{c}) Comparison between indoor and outdoor SPL. Outdoor SPL distributions with median values are shown on the top and side, respectively. (\textbf{d}) Indoor and outdoor background noise as calculated corresponding with the wind farm power output capacity $<1\%$.  }
\label{fig:results2}
\end{figure}

More AM events were detected outdoors compared to indoors, with the exception of location H3 during the daytime, as shown in \autoref{fig:results2}a. On average, outdoor AM prevalence was approximately 1.5 times higher than that measured indoors (See \autoref{fig:results2}a and supplementary Table S4). The outdoor-to-indoor AM prevalence reduction at H1 and H2 was similar, ranging between 1.5 and 2.2. In contrast, the difference between outdoor and indoor AM prevalence was smaller for data measured at H3 during the night-time (reduction = 1.1), and indoor AM occurred more often than outdoor AM during the daytime (reduction = 0.4). The AM depth measured indoors was higher than that measured outdoors (See \autoref{fig:results2}b, two sample $t$-test, all $P$ $<$ 0.001). Lower indoor background noise (or masking noise), as shown in \autoref{fig:results2}d, most likely explains the higher AM depth measured indoors. 

To examine if differences between outdoor and indoor AM prevalence could be attributed to house insulation, the distribution of simultaneously occurring outdoor and indoor noise levels are presented in \autoref{fig:results2}c. A greater A-weighted SPL reduction was observed for H1 and H2, compared to H3. This may explain some of the differences between the relative outdoor and indoor AM prevalence for H3 (\autoref{fig:results2}a). The outdoor-to-indoor SPL reduction at H3 was poor for outdoor SPL $<$ 40 dBA. It should be noted that the outdoor-to-indoor noise reduction as characterised using overall noise levels depends not only on building materials, but also the noise type and indoor background noise characteristics. The lowest level of indoor background noise measured inside H3 was higher than those in H1 and H2 (\autoref{fig:results2}d). This may affect the relationship between indoor and outdoor noise levels as shown in \autoref{fig:results2}c. Although it is not very accurate to characterise the outdoor-to-indoor reduction using overall noise levels \citep{hansen2015outdoor,thorsson2018low}, this is a simple approach and the results are easy to interpret.

\subsection{Diurnal and seasonal variability}

\begin{figure}[ht]
\begin{center}
\includegraphics[width=12 cm]{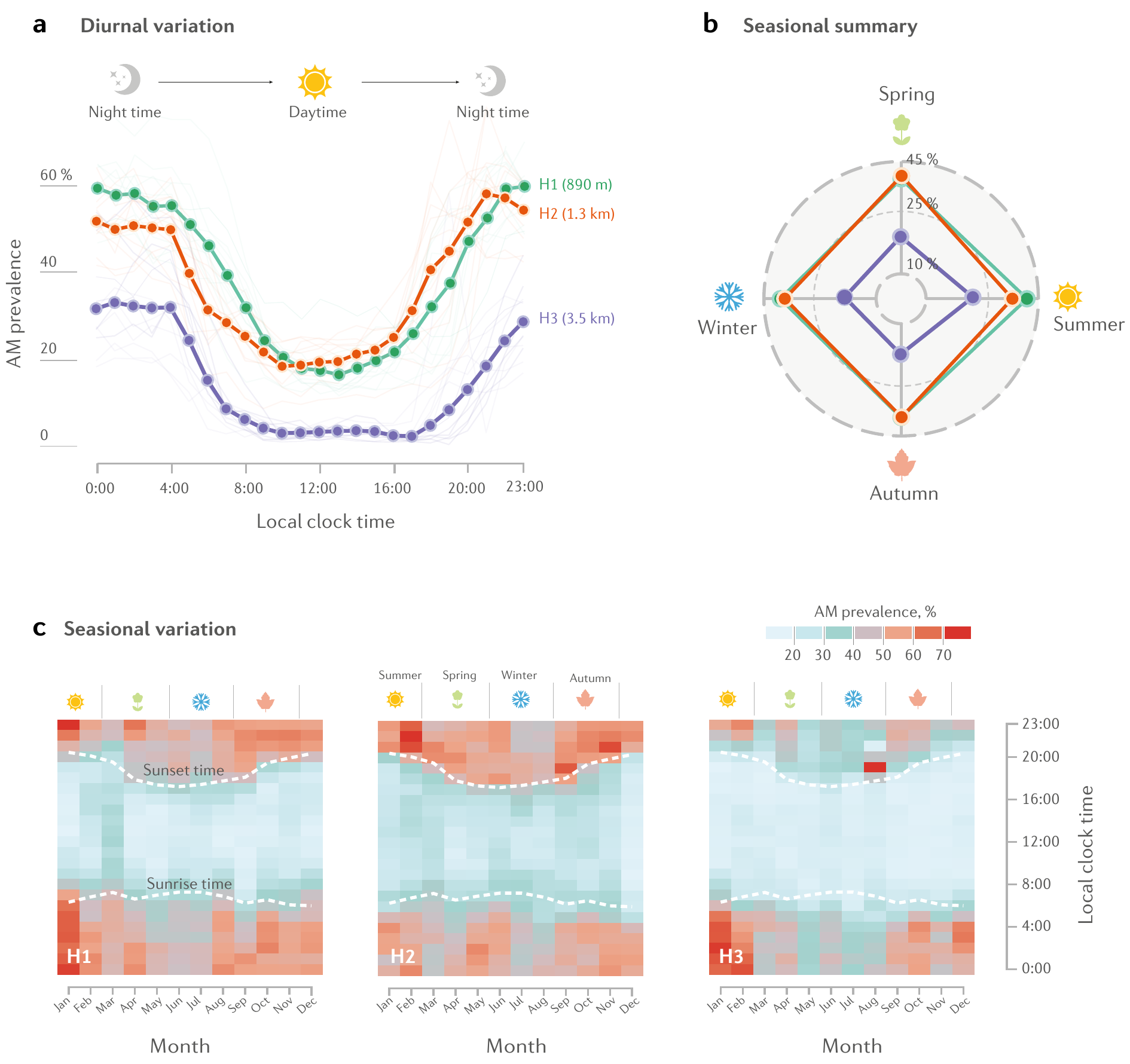}  
\end{center}
\caption{\textbf{Diurnal and seasonal variation of AM characteristics}. (\textbf{a}) Diurnal variation of AM prevalence. Thicker lines are the average trend over the year for three locations. Light lines indicate the trend for each month. (\textbf{b}) Seasonal summary of AM prevalence, calculated as the mean AM over each season (i.e., Summer from Dec-Feb, Spring from Mar-May, Winter from Jun-Aug and Autumn from Sep-Oct). (\textbf{c}) Relationship between diurnal and seasonal variation of AM prevalence. Dashed lines indicate sunset and sunrise time.}
\label{fig:result3}
\end{figure}

AM occurred most frequently at nighttime between 22:00 pm - 4:00 am, whereas the lowest AM prevalence was observed at midday around 12:00 pm (\autoref{fig:result3}a). Similar distributions of AM prevalence were observed at H1 and H3. For these locations, the highest and lowest AM prevalence were approximately 60\% and 20\%, observed at 0:00 am and 12:00 pm, respectively. For location H3, less than 5\% of AM prevalence was observed during the daytime, but this number increased to more than 30\% during the nighttime. The background noise during the nighttime was also found to be lower compared to the daytime as shown in \autoref{fig:results2}d. This was anticipated as the noise associated with human activities was expected to be lower at nighttime. Additionally, higher AM prevalence observed during the night-time could be partly attributed to lower background noise levels at nighttime compared to daytime.

The mean AM prevalence for each season is shown in \autoref{fig:result3}b. The mean AM prevalence was not notably different between months in our data. However, when AM prevalence was averaged over an hour, as shown in \autoref{fig:result3}c, clear monthly and hourly variations of AM were evident. At all measurement locations, during the winter and spring months, AM prevalence significantly increased after 16:00, which corresponds to the timing of sunset during these seasons. AM prevalence significantly increased after 20:00 in summer and autumn months, which also corresponds to the timing of sunset during these seasons. This pattern clearly corresponded to sunrise and sunset times (dashed line, \autoref{fig:result3}c) and is consistent with Larsson and \"Ohlund's \cite{Larsson2014} findings, where the authors observed a strong association between AM prevalence and solar elevation angle. 

\subsection{Relationship between meteorological and power output conditions}

\begin{figure}[ht]
\begin{center}

\includegraphics[width=13 cm]{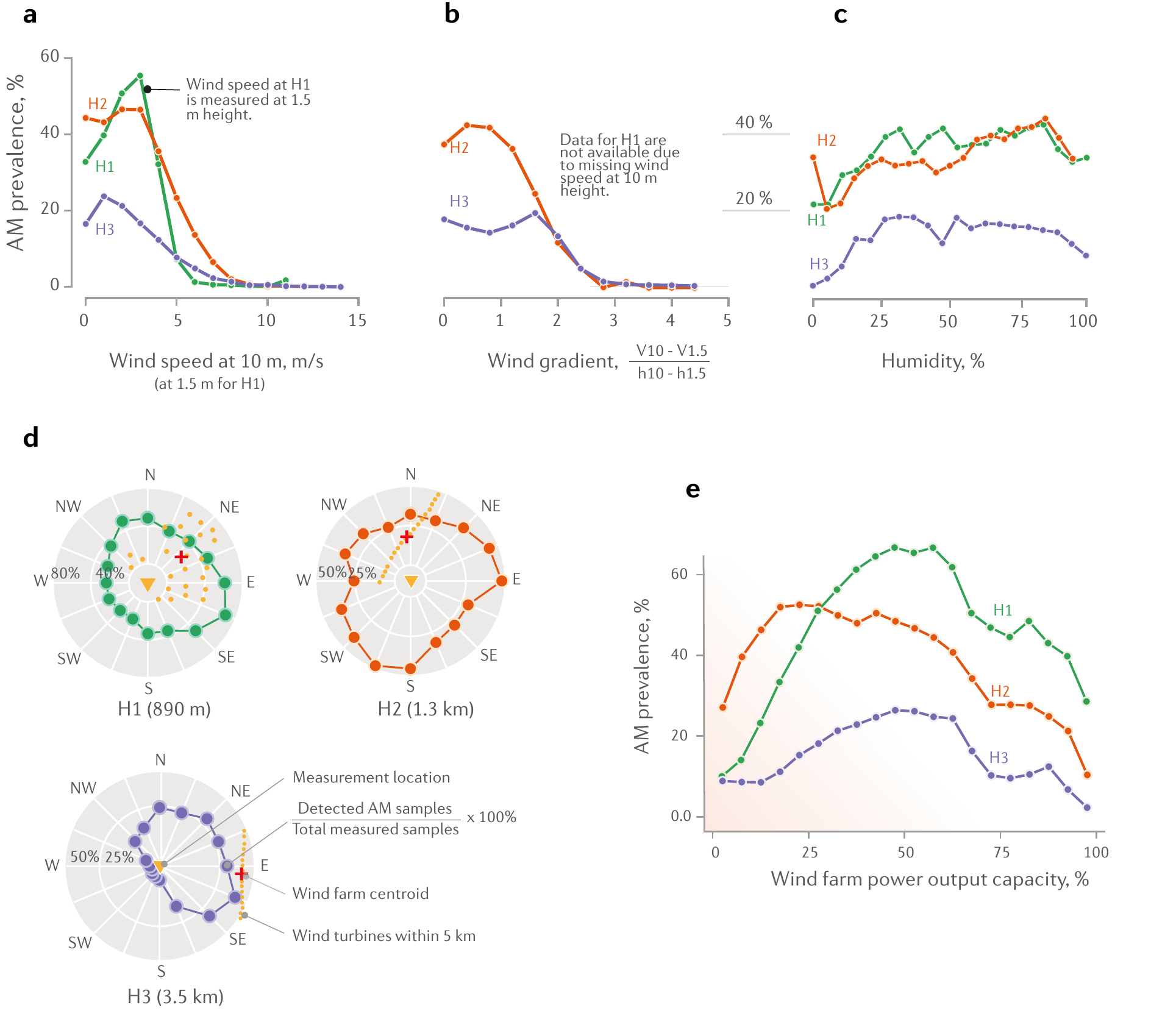}
\end{center}
\caption{\textbf{AM prevalence for different meteorological and wind farm operating conditions}. (\textbf{a,b and c}) The relationship between AM prevalence factors predicted to influence sound propagation (i.e., wind speed, wind gradient and humidity). (\textbf{d}) The dominance of AM prevalence for particular wind directions (See Supplementary Fig. S11 for measured wind speed and wind directions). The yellow dot points inside the grey circles indicate wind turbines within 5 km. The radius of the grey circles are 5 km from the measurement locations (yellow triangle points). (\textbf{e}) The relationship between wind farm percentage power output capacity and AM prevalence.}
\label{fig: figure4}
\end{figure}

An increase in the local wind speed did not always correspond to higher AM prevalence (\autoref{fig: figure4}a). The relationship between wind speed and AM prevalence at all three locations followed a similar pattern. Specifically, AM prevalence increased as the wind speed increased in specific regions (i.e., [0 4) at H1, [0 4) at H2 and [0 2) at H3), but dropped rapidly at higher wind speeds. AM was rarely detected when the local wind speed was greater than approximately 7 m/s for H1 and 10 m/s for H2 and H3. Note that these wind speed data were measured locally at each residence, rather than at the wind turbine nacelle. The wind speed data were measured at 10 m above ground level for H2 and H3 and at 1.5 m above ground level for H1, where data was not available at 10 m for the latter location.

AM prevalence measured at H2 and H3 increased with an increase of the calculated wind gradient (see \autoref{fig: figure4}b for the relevant equation) for lower values of the wind gradient (between 0 and 1 $s^{-1}$). However, for higher values of wind gradient, the prevalence of AM was reduced. AM prevalence increased with humidity (\autoref{fig: figure4}c, linear regression, $R^2$= 0.4, $P = 0.001$) as expected given that higher humidity is more favourable for sound propagation \cite{Hansen2017}. 

At all measurement locations, the AM prevalence in the upwind direction was less than for other wind directions (\autoref{fig: figure4}d, all $P < 0.001$) (See supplementary Fig. S12 to S14 for definition of wind direction category). At location H1, a mean AM prevalence of 60\% was detected for downwind data, as opposed to 30\% for upwind data. A similar trend was also observed for data measured at location H3, with around 40\% of AM events occurring during downwind conditions and less than 5\% during upwind conditions. At location H2, more AM events were detected for crosswind directions with a mean AM prevalence of around 50\% compared with 40\% downwind and 25\% in the upwind direction. 

The maximum wind farm power output did not correspond with the highest AM prevalence (\autoref{fig: figure4}e). At locations H2 and H3, AM occurred more often when the wind farm operated at around 50\% of its capacity. At location H1, AM occurred more often at low power output (between 10\% and 60\%). These findings are consistent with previous studies \cite{hansen2019prevalence}. Compared to H1 and H3, the relationship between AM prevalence and wind farm power output capacity at H2 was skewed to lower wind farm power output (\autoref{fig: figure4}e). The reasons for this observation are unclear. Possible contributions include the difference in rated turbine power output, wind farm layout, terrain characteristics or false positive detection of the algorithm. The relative importance of these factors is unclear, and thus more data and modelling approaches are needed to understand this relationship.

\newpage

\section{DISCUSSION}

This paper presented long-term AM characteristics of WFN through analysis of acoustical and meteorological data measured at three South Australian wind farms. We showed comprehensive information regarding the prevalence and diurnal distribution of AM at three locations with different wind farm layouts, wind turbine types, housing constructions and wind farm separation distances. The resulting estimates of AM depth, duration, frequency and associated sound pressure levels are important for both laboratory human trials and physical modelling of WFN \citep{makarewicz2019influence, barlas2017consistent}, as these characteristics can be used to design relevant noise stimuli and also used to validate wind farm noise models most relevant to real-world noise exposure conditions in the field. To the best of our knowledge, this is the first study to characterise and quantify AM using comprehensive data measured several kilometres from a wind farm, and to detect wind farm AM using a comprehensively validated, machine learning-based algorithm. 

AM was found to occur most often during the nighttime, consistent with previous studies \cite{hansen2019prevalence, conrady2020amplitude}. This is expected because nighttime provides favourable weather conditions for sound propagation (stable atmospheric conditions, high humidity, strong temperature inversion, high wind shear) \cite{stull2012introduction}. During these conditions, sound waves are refracted towards the ground surface in the case of downwind and crosswind conditions (although wind shear does not contribute in the latter case) \cite{ostashev2015acoustics}. This also most likely explains the high AM prevalence observed in the downwind and crosswind directions.

We found that AM is an intermittent characteristic of wind farm noise as most AM events lasted for several minutes only. This is comparable to another study \citep{Larsson2014} where the authors observed that typical main AM events lasted around 15 seconds and were followed by weaker AM. There are several factors that could lead to AM intermittency, particularly at large distances from a wind farm. These factors include varying source noise level, varying influences on noise propagation and varying background noise at the receiver location. Source noise levels vary with time as they depend on wind speed, atmospheric conditions and wake effects. Noise propagation is influenced by constantly changing atmospheric conditions. Masking noise levels at measurement locations are expected to vary with time due to changes in the local wind speed and direction and varying levels of extraneous noise. A combination of these factors could contribute towards shorter AM events observed.

The value of AM depth was highly dependent on the frequency range over which this was calculated. The largest median AM depth was associated with lower octave bands and the smallest median AM depth was associated with the variation in the overall A-weighted SPL. Given that ``swish'' noise is dominant at 500 Hz \cite{doolan2012wind} and that enhanced amplitude modulation \cite{ruk2013wind} and tonal amplitude modulation \citet{hansen2019prevalence} can occur at much lower frequencies, the overall A-weighted AM depth may be poorly correlated with human perception and is likely to substantially underestimate the perceived AM depth.  

A large difference was found between outdoor and indoor AM. At long-range, spectral imbalance of wind farm noise arises due to the higher atmospheric and ground absorption at mid to high frequencies \cite{ostashev2015acoustics}. In fact, \citet{hansen2019prevalence} found that AM usually occurs at very low frequencies (i.e., around 50 Hz) at several kilometres from a wind farm. In addition, low-frequency noise is poorly attenuated by building structures, resulting in lower outdoor-to-indoor noise level reduction at low frequencies \cite{hansen2015outdoor}. These results could explain the relatively small outdoor-to-indoor reduction in AM prevalence that was observed at H3 at nighttime. The increase in AM events measured indoors during the daytime at H3 may have been a result of high outdoor ambient noise that masked the outdoor AM but not the indoor AM. These findings suggest that the outdoor-to-indoor noise reduction also impacts AM prevalence. Also, a greater AM depth is associated with higher annoyance \cite{Lee2011,Schaffer2016,yokoyama2013study}, and thus AM may be more annoying when people are indoors with low ambient background noise, which is exaggerated during the nighttime. These observations are particularly relevant for cases where AM is only measured outdoors.

We investigated the seasonal variations of AM prevalence measured outdoors, but found no significant differences in outdoor AM prevalence between seasons. This contrasts with the study conducted by \citep{conrady2020amplitude} where the authors reported more frequent AM during the Winter compared to spring, but with more limited data from a much colder climate in Sweden. Interestingly, we found a remarkably strong temporal relationship between sunset and sunrise times and the beginning and end of AM, most likely indicative of temperature inversion effects \cite{stull2012introduction}.

While directivity of broadband trailing edge noise causes swishing noise which is prominent close to the turbine (within 1-2 rotor diameters) \citep{oerlemans2015effect}, the wind gradient has been hypothesised to cause AM perceived at larger distances (hundreds of metres to several km) from the wind farm. This is due to an increased difference in aerodynamic loading between the upper and lower parts of the wind turbine blade trajectory \cite{bowdler2008amplitude,vandenberg2005beat,oerlemans2015effect}. This change in loading could then affect blade aerodynamic noise production such as trailing edge, leading edge and loading noise sources which would show greatest variation between the lowest and highest parts of the blade trajectories, thereby resulting in AM. It is thus expected that higher wind gradients provide more favourable conditions for AM generation and that AM prevalence increases with increasing wind gradient. However, our data did not support this as AM prevalence was reduced with increasing local wind gradient, a finding comparable with other studies \cite{conrady2020amplitude,cooper2013automated}. However, it is also worth noting that the wind gradient measured in our study was based on local wind speeds between 1.5 and 10 m which is likely not representative of wind gradients at higher altitudes more relevant for AM generation at the noise source. Furthermore, wind gradients over a ridge are significantly modified by wind speed-up effects \cite{ngo2009experimental}. Therefore, the association between wind shear and AM is still unclear and more suitable wind speed data closer to the noise source are needed to confirm the wind gradient hypothesis.

Although we measured comprehensive acoustical data, a limitation of our study is a lack of comprehensive meteorological data measured at hub-height most relevant to the noise source. Thus, relationships between AM and meteorological conditions remain unclear and further studies are needed to more comprehensively assess these relationships. This limitation calls for better data sharing practices between wind farm operators and researchers \cite{kusiak2016renewables} to allow for more in depth analysis of relationships between wind farm noise and meteorological conditions.

\section{CONCLUSION}

In summary, this study characterised and quantified wind farm noise AM for a large data set measured over one year at three relatively long-range distances from three wind farms in South Australia. At nighttime, AM prevalence was lower indoors than outdoors, but there was an increase in AM depth in the indoor data. Our findings also showed a dependence of AM prevalence with respect to time (i.e., diurnal and monthly variations). We further found that AM occurred more often during downwind and crosswind directions, compared to upwind conditions. The measured data can also be used for validating wind farm propagation models, particularly those that attempt to model AM. Ultimately, improved wind farm noise assessment guidelines and more accurate noise prediction models will make wind energy more acceptable to surrounding communities.

\section*{Acknowledgement}
The authors gratefully acknowledge financial support from the Australian Research Council, projects DP120102185 and DE180100022 and the National Health and Medical Research Council, Project 1113571. The author PDN was supported by a Flinders University Research Scholarship (FURS) for this work.




\bibliographystyle{elsarticle-harv} 
\bibliography{WTN_all.bib}

\begin{thebibliography}{48}
\expandafter\ifx\csname natexlab\endcsname\relax\def\natexlab#1{#1}\fi
\providecommand{\url}[1]{\texttt{#1}}
\providecommand{\href}[2]{#2}
\providecommand{\path}[1]{#1}
\providecommand{\DOIprefix}{doi:}
\providecommand{\ArXivprefix}{arXiv:}
\providecommand{\URLprefix}{URL: }
\providecommand{\Pubmedprefix}{pmid:}
\providecommand{\doi}[1]{\href{http://dx.doi.org/#1}{\path{#1}}}
\providecommand{\Pubmed}[1]{\href{pmid:#1}{\path{#1}}}
\providecommand{\bibinfo}[2]{#2}
\ifx\xfnm\relax \def\xfnm[#1]{\unskip,\space#1}\fi
\bibitem[{Barlas et~al.(2017)Barlas, Zhu, Shen, Dag and
  Moriarty}]{barlas2017consistent}
\bibinfo{author}{Barlas, E.}, \bibinfo{author}{Zhu, W.J.},
  \bibinfo{author}{Shen, W.Z.}, \bibinfo{author}{Dag, K.O.},
  \bibinfo{author}{Moriarty, P.}, \bibinfo{year}{2017}.
\newblock \bibinfo{title}{Consistent modelling of wind turbine noise
  propagation from source to receiver}.
\newblock \bibinfo{journal}{The Journal of the Acoustical Society of America}
  \bibinfo{volume}{142}, \bibinfo{pages}{3297--3310}.
\bibitem[{Bass et~al.(2016)Bass, Cand, Coles, Davis, Irvine, Leventhall, Levet,
  Miller, Sexton and Shelton}]{Bass2016}
\bibinfo{author}{Bass, J.}, \bibinfo{author}{Cand, M.}, \bibinfo{author}{Coles,
  D.}, \bibinfo{author}{Davis, R.}, \bibinfo{author}{Irvine, G.},
  \bibinfo{author}{Leventhall, G.}, \bibinfo{author}{Levet, T.},
  \bibinfo{author}{Miller, S.}, \bibinfo{author}{Sexton, D.},
  \bibinfo{author}{Shelton, J.}, \bibinfo{year}{2016}.
\newblock \bibinfo{title}{Institute of acoustics ioa noise working group ( wind
  turbine noise ) amplitude modulation working group final report a method for
  rating amplitude modulation in wind turbine noise version 1}.
\newblock \bibinfo{journal}{IOA report} .
\bibitem[{Van~den Berg(2005)}]{vandenberg2005beat}
\bibinfo{author}{Van~den Berg, G.}, \bibinfo{year}{2005}.
\newblock \bibinfo{title}{The beat is getting stronger: the effect of
  atmospheric stability on low frequency modulated sound of wind turbines}.
\newblock \bibinfo{journal}{Journal of Low Frequency Noise, Vibration and
  Active Control} \bibinfo{volume}{24}, \bibinfo{pages}{1--24}.
\bibitem[{Bies et~al.(2017)Bies, Hansen and Howard}]{bies2017engineering}
\bibinfo{author}{Bies, D.A.}, \bibinfo{author}{Hansen, C.},
  \bibinfo{author}{Howard, C.}, \bibinfo{year}{2017}.
\newblock \bibinfo{title}{Engineering noise control}.
\newblock \bibinfo{publisher}{CRC press}.
\bibitem[{Bowdler(2008)}]{bowdler2008amplitude}
\bibinfo{author}{Bowdler, D.}, \bibinfo{year}{2008}.
\newblock \bibinfo{title}{Amplitude modulation of wind turbine noise: a review
  of the evidence}.
\newblock \bibinfo{journal}{Institute of Acoustics Bulletin}
  \bibinfo{volume}{33}, \bibinfo{pages}{31--41}.
\bibitem[{Chawla et~al.(2003)Chawla, Lazarevic, Hall and
  Bowyer}]{chawla2003smoteboost}
\bibinfo{author}{Chawla, N.V.}, \bibinfo{author}{Lazarevic, A.},
  \bibinfo{author}{Hall, L.O.}, \bibinfo{author}{Bowyer, K.W.},
  \bibinfo{year}{2003}.
\newblock \bibinfo{title}{Smoteboost: Improving prediction of the minority
  class in boosting}, in: \bibinfo{booktitle}{European conference on principles
  of data mining and knowledge discovery}, \bibinfo{organization}{Springer}.
  pp. \bibinfo{pages}{107--119}.
\bibitem[{Conrady et~al.(2020)Conrady, Bolin, Sj{\"o}blom and
  Rutgersson}]{conrady2020amplitude}
\bibinfo{author}{Conrady, K.}, \bibinfo{author}{Bolin, K.},
  \bibinfo{author}{Sj{\"o}blom, A.}, \bibinfo{author}{Rutgersson, A.},
  \bibinfo{year}{2020}.
\newblock \bibinfo{title}{Amplitude modulation of wind turbine sound in cold
  climates}.
\newblock \bibinfo{journal}{Applied Acoustics} \bibinfo{volume}{158},
  \bibinfo{pages}{107024}.
\bibitem[{Cooper and Evans(2013)}]{cooper2013automated}
\bibinfo{author}{Cooper, J.}, \bibinfo{author}{Evans, T.},
  \bibinfo{year}{2013}.
\newblock \bibinfo{title}{Automated detection and analysis of amplitude
  modulation at a residence and wind turbine}.
\newblock \bibinfo{journal}{Proceedings of Acoustics 2013} .
\bibitem[{Doolan et~al.(2012)Doolan, Moreau and Brooks}]{doolan2012wind}
\bibinfo{author}{Doolan, C.J.}, \bibinfo{author}{Moreau, D.J.},
  \bibinfo{author}{Brooks, L.A.}, \bibinfo{year}{2012}.
\newblock \bibinfo{title}{Wind turbine noise mechanisms and some concepts for
  its control.}
\newblock \bibinfo{journal}{Acoustics Australia} \bibinfo{volume}{40}.
\bibitem[{GWEC(2019)}]{gwec2019global}
\bibinfo{author}{GWEC, G.W.E.C.}, \bibinfo{year}{2019}.
\newblock \bibinfo{title}{Global wind report: Annual market update 2019}.
\newblock \bibinfo{journal}{URL http://gwec.
  net/global-figures/graphs/.[Accessed Feb 14, 2021]} .
\bibitem[{Hansen et~al.(2017)Hansen, Doolan and Hansen}]{Hansen2017}
\bibinfo{author}{Hansen, C.H.}, \bibinfo{author}{Doolan, C.J.},
  \bibinfo{author}{Hansen, K.L.}, \bibinfo{year}{2017}.
\newblock \bibinfo{title}{Wind Farm Noise: Measurement, Assessment and
  Control}.
\newblock \bibinfo{edition}{1} ed., \bibinfo{publisher}{John Wiley \& Sons
  Ltd}.
\bibitem[{Hansen et~al.(2015)Hansen, C. and Zajamsek}]{hansen2015outdoor}
\bibinfo{author}{Hansen, K.}, \bibinfo{author}{C., H.},
  \bibinfo{author}{Zajamsek, B.}, \bibinfo{year}{2015}.
\newblock \bibinfo{title}{Outdoor to indoor reduction of wind farm noise for
  rural residences}.
\newblock \bibinfo{journal}{Building and Environment} \bibinfo{volume}{94},
  \bibinfo{pages}{764--772}.
\bibitem[{Hansen et~al.(2014a)Hansen, Zajamsek and Hansen}]{hansen2014comp}
\bibinfo{author}{Hansen, K.}, \bibinfo{author}{Zajamsek, B.},
  \bibinfo{author}{Hansen, C.}, \bibinfo{year}{2014}a.
\newblock \bibinfo{title}{Comparison of the noise levels measured in the
  vicinity of a wind farm for shutdown and operational conditions}, in:
  \bibinfo{booktitle}{INTER-NOISE and NOISE-CON congress and conference
  proceedings}, \bibinfo{organization}{Institute of Noise Control Engineering}.
  pp. \bibinfo{pages}{5192--5202}.
\bibitem[{Hansen et~al.(2014b)Hansen, Zajamsek and
  Hansen}]{hansen2014identification}
\bibinfo{author}{Hansen, K.}, \bibinfo{author}{Zajamsek, B.},
  \bibinfo{author}{Hansen, C.}, \bibinfo{year}{2014}b.
\newblock \bibinfo{title}{Identification of low frequency wind turbine noise
  using secondary windscreens of various geometries}.
\newblock \bibinfo{journal}{Noise Control Engineering Journal}
  \bibinfo{volume}{62}, \bibinfo{pages}{69--82}.
\bibitem[{Hansen et~al.(2019)Hansen, Nguyen, Zajam{\v{s}}ek, Catcheside and
  Hansen}]{hansen2019prevalence}
\bibinfo{author}{Hansen, K.L.}, \bibinfo{author}{Nguyen, P.},
  \bibinfo{author}{Zajam{\v{s}}ek, B.}, \bibinfo{author}{Catcheside, P.},
  \bibinfo{author}{Hansen, C.H.}, \bibinfo{year}{2019}.
\newblock \bibinfo{title}{Prevalence of wind farm amplitude modulation at
  long-range residential locations}.
\newblock \bibinfo{journal}{Journal of Sound and Vibration}
  \bibinfo{volume}{455}, \bibinfo{pages}{136--149}.
\bibitem[{Ingielewicz et~al.(2014)Ingielewicz, Zagubie{\'n}
  et~al.}]{ingielewicz2014infrasound}
\bibinfo{author}{Ingielewicz, R.}, \bibinfo{author}{Zagubie{\'n}, A.}, et~al.,
  \bibinfo{year}{2014}.
\newblock \bibinfo{title}{Infrasound noise of natural sources in environment
  and infrasound noise of wind turbines}.
\newblock \bibinfo{journal}{Polish Journal of Environmental Studies}
  \bibinfo{volume}{23}, \bibinfo{pages}{1323--1327}.
\bibitem[{Ioannidou et~al.(2016)Ioannidou, Santurette and
  Jeong}]{Ioannidou2016}
\bibinfo{author}{Ioannidou, C.}, \bibinfo{author}{Santurette, S.},
  \bibinfo{author}{Jeong, C.H.}, \bibinfo{year}{2016}.
\newblock \bibinfo{title}{Effect of modulation depth, frequency, and
  intermittence on wind turbine noise annoyance}.
\newblock \bibinfo{journal}{The Journal of the Acoustical Society of America}
  \bibinfo{volume}{139}, \bibinfo{pages}{1241--1251}.
\newblock \URLprefix \url{http://asa.scitation.org/doi/10.1121/1.4944570},
  \DOIprefix\doi{10.1121/1.4944570}.
\bibitem[{Jurado et~al.(2019)Jurado, Gordillo and Moore}]{jurado2019effect}
\bibinfo{author}{Jurado, C.}, \bibinfo{author}{Gordillo, D.},
  \bibinfo{author}{Moore, B.C.}, \bibinfo{year}{2019}.
\newblock \bibinfo{title}{Effect of amplitude fluctuations on the loudness of
  low-frequency sounds}.
\newblock \bibinfo{journal}{The Journal of the Acoustical Society of America}
  \bibinfo{volume}{146}, \bibinfo{pages}{3047--3047}.
\bibitem[{Kitzing et~al.(2020)Kitzing, Jensen, Telsnig and
  Lantz}]{kitzing2020multifaceted}
\bibinfo{author}{Kitzing, L.}, \bibinfo{author}{Jensen, M.K.},
  \bibinfo{author}{Telsnig, T.}, \bibinfo{author}{Lantz, E.},
  \bibinfo{year}{2020}.
\newblock \bibinfo{title}{Multifaceted drivers for onshore wind energy
  repowering and their implications for energy transition}.
\newblock \bibinfo{journal}{Nature Energy} , \bibinfo{pages}{1--10}.
\bibitem[{Krohn and Damborg(1999)}]{krohn1999public}
\bibinfo{author}{Krohn, S.}, \bibinfo{author}{Damborg, S.},
  \bibinfo{year}{1999}.
\newblock \bibinfo{title}{On public attitudes towards wind power}.
\newblock \bibinfo{journal}{Renewable energy} \bibinfo{volume}{16},
  \bibinfo{pages}{954--960}.
\bibitem[{Kusiak(2016)}]{kusiak2016renewables}
\bibinfo{author}{Kusiak, A.}, \bibinfo{year}{2016}.
\newblock \bibinfo{title}{Renewables: Share data on wind energy}.
\newblock \bibinfo{journal}{Nature} \bibinfo{volume}{529},
  \bibinfo{pages}{19--21}.
\bibitem[{Larsson and \"Ohlund(2014)}]{Larsson2014}
\bibinfo{author}{Larsson, C.}, \bibinfo{author}{\"Ohlund, O.},
  \bibinfo{year}{2014}.
\newblock \bibinfo{title}{Amplitude modulation of sound from wind turbines
  under various meteorological conditions}.
\newblock \bibinfo{journal}{Journal of the Acoustical Society of America}
  \bibinfo{volume}{135}, \bibinfo{pages}{67--73}.
\bibitem[{Lee et~al.(2011)Lee, Kim, Choi and Lee}]{Lee2011}
\bibinfo{author}{Lee, S.}, \bibinfo{author}{Kim, K.}, \bibinfo{author}{Choi,
  W.}, \bibinfo{author}{Lee, S.}, \bibinfo{year}{2011}.
\newblock \bibinfo{title}{Annoyance caused by amplitude modulation of wind
  turbine noise}.
\newblock \bibinfo{journal}{Noise Control Engineering Journal}
  \bibinfo{volume}{59}, \bibinfo{pages}{38}.
\newblock \DOIprefix\doi{10.3397/1.3531797}.
\bibitem[{Liebich et~al.(2020)Liebich, Lack, Hansen, Zajam{\v{s}}ek, Lovato,
  Catcheside and Micic}]{liebich2020systematic}
\bibinfo{author}{Liebich, T.}, \bibinfo{author}{Lack, L.},
  \bibinfo{author}{Hansen, K.}, \bibinfo{author}{Zajam{\v{s}}ek, B.},
  \bibinfo{author}{Lovato, N.}, \bibinfo{author}{Catcheside, P.},
  \bibinfo{author}{Micic, G.}, \bibinfo{year}{2020}.
\newblock \bibinfo{title}{A systematic review and meta-analysis of wind turbine
  noise effects on sleep using validated objective and subjective sleep
  assessments}.
\newblock \bibinfo{journal}{Journal of Sleep Research} ,
  \bibinfo{pages}{e13228}.
\bibitem[{Liu et~al.(2012)Liu, Bo and Veidt}]{liu2012tonality}
\bibinfo{author}{Liu, X.}, \bibinfo{author}{Bo, L.}, \bibinfo{author}{Veidt,
  M.}, \bibinfo{year}{2012}.
\newblock \bibinfo{title}{Tonality evaluation of wind turbine noise by
  filter-segmentation}.
\newblock \bibinfo{journal}{Measurement} \bibinfo{volume}{45},
  \bibinfo{pages}{711--718}.
\bibitem[{Makarewicz and Go{\l}ebiewski(2019)}]{makarewicz2019influence}
\bibinfo{author}{Makarewicz, R.}, \bibinfo{author}{Go{\l}ebiewski, R.},
  \bibinfo{year}{2019}.
\newblock \bibinfo{title}{The influence of a low level jet on the thumps
  generated by a wind turbine}.
\newblock \bibinfo{journal}{Renewable and Sustainable Energy Reviews}
  \bibinfo{volume}{104}, \bibinfo{pages}{337--342}.
\bibitem[{Micic et~al.(2018)Micic, Zajamsek, Lack, Hansen, Doolan, Hansen,
  Vakulin, Lovato, Bruck, Chai-Coetzer et~al.}]{Micic2018}
\bibinfo{author}{Micic, G.}, \bibinfo{author}{Zajamsek, B.},
  \bibinfo{author}{Lack, L.}, \bibinfo{author}{Hansen, K.},
  \bibinfo{author}{Doolan, C.}, \bibinfo{author}{Hansen, C.},
  \bibinfo{author}{Vakulin, A.}, \bibinfo{author}{Lovato, N.},
  \bibinfo{author}{Bruck, D.}, \bibinfo{author}{Chai-Coetzer, C.L.}, et~al.,
  \bibinfo{year}{2018}.
\newblock \bibinfo{title}{A review of the potential impacts of wind farm noise
  on sleep}.
\newblock \bibinfo{journal}{Acoustics Australia} \bibinfo{volume}{46},
  \bibinfo{pages}{87--97}.
\bibitem[{Ngo and Letchford(2009)}]{ngo2009experimental}
\bibinfo{author}{Ngo, T.T.}, \bibinfo{author}{Letchford, C.W.},
  \bibinfo{year}{2009}.
\newblock \bibinfo{title}{Experimental study of topographic effects on gust
  wind speed}.
\newblock \bibinfo{journal}{Journal of wind engineering and industrial
  aerodynamics} \bibinfo{volume}{97}, \bibinfo{pages}{426--438}.
\bibitem[{Oerlemans(2015)}]{oerlemans2015effect}
\bibinfo{author}{Oerlemans, S.}, \bibinfo{year}{2015}.
\newblock \bibinfo{title}{Effect of wind shear on amplitude modulation of wind
  turbine noise}.
\newblock \bibinfo{journal}{International Journal of Aeroacoustics}
  \bibinfo{volume}{14}, \bibinfo{pages}{715--728}.
\bibitem[{Ostashev and Wilson(2015)}]{ostashev2015acoustics}
\bibinfo{author}{Ostashev, V.E.}, \bibinfo{author}{Wilson, D.K.},
  \bibinfo{year}{2015}.
\newblock \bibinfo{title}{Acoustics in moving inhomogeneous media}.
\newblock \bibinfo{publisher}{CRC Press}.
\bibitem[{Paulraj and V{\"a}lisuo(2017)}]{paulraj2017effect}
\bibinfo{author}{Paulraj, T.}, \bibinfo{author}{V{\"a}lisuo, P.},
  \bibinfo{year}{2017}.
\newblock \bibinfo{title}{Effect of wind speed and wind direction on amplitude
  modulation of wind turbine noise}, in: \bibinfo{booktitle}{INTER-NOISE and
  NOISE-CON Congress and Conference Proceedings},
  \bibinfo{organization}{Institute of Noise Control Engineering}. pp.
  \bibinfo{pages}{5479--5489}.
\bibitem[{Perkins et~al.(2016)Perkins, Lotinga, Berry, Grimwood and
  Stansfeld}]{perkins2016review}
\bibinfo{author}{Perkins, R.}, \bibinfo{author}{Lotinga, M.},
  \bibinfo{author}{Berry, B.}, \bibinfo{author}{Grimwood, C.},
  \bibinfo{author}{Stansfeld, S.}, \bibinfo{year}{2016}.
\newblock \bibinfo{title}{A review of research into the human response to
  amplitude modulated wind turbine noise and development of a planning control
  method}, in: \bibinfo{booktitle}{INTER-NOISE and NOISE-CON Congress and
  Conference Proceedings}, \bibinfo{organization}{Institute of Noise Control
  Engineering}. pp. \bibinfo{pages}{5222--5233}.
\bibitem[{Sch\"affer et~al.(2016)Sch\"affer, Schlittmeier, Pieren, Heutschi,
  Brink, Graf and Hellbr\"uck}]{Schaffer2016}
\bibinfo{author}{Sch\"affer, B.}, \bibinfo{author}{Schlittmeier, S.J.},
  \bibinfo{author}{Pieren, R.}, \bibinfo{author}{Heutschi, K.},
  \bibinfo{author}{Brink, M.}, \bibinfo{author}{Graf, R.},
  \bibinfo{author}{Hellbr\"uck, J.}, \bibinfo{year}{2016}.
\newblock \bibinfo{title}{Short-term annoyance reactions to stationary and
  time-varying wind turbine and road traffic noise: A laboratory study}.
\newblock \bibinfo{journal}{The Journal of the Acoustical Society of America}
  \bibinfo{volume}{139}, \bibinfo{pages}{2949--2963}.
\newblock \URLprefix \url{http://asa.scitation.org/doi/10.1121/1.4949566},
  \DOIprefix\doi{10.1121/1.4949566}.
\bibitem[{Schuster et~al.(2015)Schuster, Bulling and
  K{\"o}ppel}]{schuster2015consolidating}
\bibinfo{author}{Schuster, E.}, \bibinfo{author}{Bulling, L.},
  \bibinfo{author}{K{\"o}ppel, J.}, \bibinfo{year}{2015}.
\newblock \bibinfo{title}{Consolidating the state of knowledge: a synoptical
  review of wind energy’s wildlife effects}.
\newblock \bibinfo{journal}{Environmental management} \bibinfo{volume}{56},
  \bibinfo{pages}{300--331}.
\bibitem[{Seiffert et~al.(2009)Seiffert, Khoshgoftaar, Van~Hulse and
  Napolitano}]{seiffert2009rusboost}
\bibinfo{author}{Seiffert, C.}, \bibinfo{author}{Khoshgoftaar, T.M.},
  \bibinfo{author}{Van~Hulse, J.}, \bibinfo{author}{Napolitano, A.},
  \bibinfo{year}{2009}.
\newblock \bibinfo{title}{Rusboost: A hybrid approach to alleviating class
  imbalance}.
\newblock \bibinfo{journal}{IEEE Transactions on Systems, Man, and
  Cybernetics-Part A: Systems and Humans} \bibinfo{volume}{40},
  \bibinfo{pages}{185--197}.
\bibitem[{Smith et~al.(2020)Smith, {\"O}gren, Thorsson, Hussain-Alkhateeb,
  Pedersen, Forss{\'e}n, Ageborg~Morsing and
  Persson~Waye}]{smith2020laboratory}
\bibinfo{author}{Smith, M.G.}, \bibinfo{author}{{\"O}gren, M.},
  \bibinfo{author}{Thorsson, P.}, \bibinfo{author}{Hussain-Alkhateeb, L.},
  \bibinfo{author}{Pedersen, E.}, \bibinfo{author}{Forss{\'e}n, J.},
  \bibinfo{author}{Ageborg~Morsing, J.}, \bibinfo{author}{Persson~Waye, K.},
  \bibinfo{year}{2020}.
\newblock \bibinfo{title}{A laboratory study on the effects of wind turbine
  noise on sleep: results of the polysomnographic witnes study}.
\newblock \bibinfo{journal}{Sleep} .
\bibitem[{Stull(2012)}]{stull2012introduction}
\bibinfo{author}{Stull, R.B.}, \bibinfo{year}{2012}.
\newblock \bibinfo{title}{An introduction to boundary layer meteorology}.
  volume~\bibinfo{volume}{13}.
\newblock \bibinfo{publisher}{Springer Science \& Business Media}.
\bibitem[{Thaker et~al.(2018)Thaker, Zambre and Bhosale}]{thaker2018wind}
\bibinfo{author}{Thaker, M.}, \bibinfo{author}{Zambre, A.},
  \bibinfo{author}{Bhosale, H.}, \bibinfo{year}{2018}.
\newblock \bibinfo{title}{Wind farms have cascading impacts on ecosystems
  across trophic levels}.
\newblock \bibinfo{journal}{Nature ecology \& evolution} \bibinfo{volume}{2},
  \bibinfo{pages}{1854--1858}.
\bibitem[{Thorsson et~al.(2018)Thorsson, Persson~Waye, Smith, {\"O}gren,
  Pedersen and Forss{\'e}n}]{thorsson2018low}
\bibinfo{author}{Thorsson, P.}, \bibinfo{author}{Persson~Waye, K.},
  \bibinfo{author}{Smith, M.}, \bibinfo{author}{{\"O}gren, M.},
  \bibinfo{author}{Pedersen, E.}, \bibinfo{author}{Forss{\'e}n, J.},
  \bibinfo{year}{2018}.
\newblock \bibinfo{title}{Low-frequency outdoor--indoor noise level difference
  for wind turbine assessment}.
\newblock \bibinfo{journal}{The Journal of the Acoustical Society of America}
  \bibinfo{volume}{143}, \bibinfo{pages}{EL206--EL211}.
\bibitem[{UK(2013)}]{ruk2013wind}
\bibinfo{author}{UK, R.}, \bibinfo{year}{2013}.
\newblock \bibinfo{title}{Wind turbine amplitude modulation: research to
  improve understanding as to its cause \& effect}.
\newblock \bibinfo{type}{Technical Report}. Renewable UK.
\bibitem[{Vautard et~al.(2014)Vautard, Thais, Tobin, Br{\'e}on, De~Lavergne,
  Colette, Yiou and Ruti}]{vautard2014regional}
\bibinfo{author}{Vautard, R.}, \bibinfo{author}{Thais, F.},
  \bibinfo{author}{Tobin, I.}, \bibinfo{author}{Br{\'e}on, F.M.},
  \bibinfo{author}{De~Lavergne, J.g.D.}, \bibinfo{author}{Colette, A.},
  \bibinfo{author}{Yiou, P.}, \bibinfo{author}{Ruti, P.M.},
  \bibinfo{year}{2014}.
\newblock \bibinfo{title}{Regional climate model simulations indicate limited
  climatic impacts by operational and planned european wind farms}.
\newblock \bibinfo{journal}{Nature communications} \bibinfo{volume}{5},
  \bibinfo{pages}{3196}.
\bibitem[{Von~H{\"u}nerbein et~al.(2013)Von~H{\"u}nerbein, King, Piper and
  Cand}]{von2013wind}
\bibinfo{author}{Von~H{\"u}nerbein, S.}, \bibinfo{author}{King, A.},
  \bibinfo{author}{Piper, B.}, \bibinfo{author}{Cand, M.},
  \bibinfo{year}{2013}.
\newblock \bibinfo{title}{Wind turbine amplitude modulation: Research to
  improve understanding as to its cause and effect--work package b (2):
  Development of an am dose-response relationship}.
\newblock \bibinfo{journal}{RenewableUK, http://www. renewableuk. com/(Last
  viewed April 11, 2016)} .
\bibitem[{WindPower(2021)}]{windpower2021}
\bibinfo{author}{WindPower, T.W.P.}, \bibinfo{year}{2021}.
\newblock \bibinfo{title}{Wind farms databases}.
\newblock \bibinfo{journal}{URL https://www.thewindpower.net/. [Accessed Feb
  11, 2021]} .
\bibitem[{Wolsink(2007)}]{wolsink2007wind}
\bibinfo{author}{Wolsink, M.}, \bibinfo{year}{2007}.
\newblock \bibinfo{title}{Wind power implementation: the nature of public
  attitudes: equity and fairness instead of ‘backyard motives’}.
\newblock \bibinfo{journal}{Renewable and sustainable energy reviews}
  \bibinfo{volume}{11}, \bibinfo{pages}{1188--1207}.
\bibitem[{Yokoyama et~al.(2013)Yokoyama, Sakamoto and
  Tachibana}]{yokoyama2013study}
\bibinfo{author}{Yokoyama, S.}, \bibinfo{author}{Sakamoto, S.},
  \bibinfo{author}{Tachibana, H.}, \bibinfo{year}{2013}.
\newblock \bibinfo{title}{Study on the amplitude modulation of wind turbine
  noise: part 2-auditory experiments}, in: \bibinfo{booktitle}{INTER-NOISE and
  NOISE-CON Congress and Conference Proceedings},
  \bibinfo{organization}{Institute of Noise Control Engineering}. pp.
  \bibinfo{pages}{3136--3145}.
\bibitem[{Zagubie{\'n}(2018)}]{zagubien2018analysis}
\bibinfo{author}{Zagubie{\'n}, A.}, \bibinfo{year}{2018}.
\newblock \bibinfo{title}{Analysis of acoustic pressure fluctuation around wind
  farms.}
\newblock \bibinfo{journal}{Polish Journal of Environmental Studies}
  \bibinfo{volume}{27}.
\bibitem[{Zajamsek et~al.(2016)Zajamsek, Hansen, Doolan and
  Hansen}]{Zajamsek2016}
\bibinfo{author}{Zajamsek, B.}, \bibinfo{author}{Hansen, K.L.},
  \bibinfo{author}{Doolan, C.J.}, \bibinfo{author}{Hansen, C.H.},
  \bibinfo{year}{2016}.
\newblock \bibinfo{title}{Characterisation of wind farm infrasound and
  low-frequency noise}.
\newblock \bibinfo{journal}{Journal of Sound and Vibration} ,
  \bibinfo{pages}{1--15}\DOIprefix\doi{10.1016/j.jsv.2016.02.001}.
\bibitem[{Zhou et~al.(2012)Zhou, Tian, Roy, Thorncroft, Bosart and
  Hu}]{zhou2012impacts}
\bibinfo{author}{Zhou, L.}, \bibinfo{author}{Tian, Y.}, \bibinfo{author}{Roy,
  S.B.}, \bibinfo{author}{Thorncroft, C.}, \bibinfo{author}{Bosart, L.F.},
  \bibinfo{author}{Hu, Y.}, \bibinfo{year}{2012}.
\newblock \bibinfo{title}{Impacts of wind farms on land surface temperature}.
\newblock \bibinfo{journal}{Nature Climate Change} \bibinfo{volume}{2},
  \bibinfo{pages}{539--543}.

\end{thebibliography}






\end{document}